\documentclass[aps,prl,twocolumn,showpacs,amsmath,amssymb]{revtex4}
\usepackage{epsfig}
\usepackage{bm}

\begin{document}
\title{Proposal for non-local electron-hole turnstile in the Quantum Hall regime.}

\author{F. Battista and P. Samuelsson} \affiliation{Division of
Mathematical Physics, Lund University, Box 118, S-221 00 Lund, Sweden}

\begin{abstract} 
  We present a theory for a mesoscopic turnstile that produces
  spatially separated streams of electrons and holes along edge states
  in the quantum Hall regime. For a broad range of frequencies in the
  non-adiabatic regime the turnstile operation is found to be ideal,
  producing one electron and one hole per cycle. The accuracy of the
  turnstile operation is characterized by the fluctuations of the
  transferred charge per cycle. The fluctuations are found to be
  negligibly small in the ideal regime.
\end{abstract}

\maketitle 

Transport along edge states in the integer quantum Hall regime has
recently attracted large interest. The unidirectional transport
properties of the edge states together with the possibility of using
quantum point contacts as beam splitters has motivated a number of
experiments on electronic analogues of optical interferometers, such
as single particle Mach Zehnder \cite{MZI} and two particle Hanbury
Brown Twiss \cite{2PI} interferometers. In the experiment by Altimiras
{\it et al} \cite{pierre} the electronic-optic analogue was supported
by probing the non-equilibrium electronic distribution along the
edge. Moreover, the prospect of entanglement generation in electronic
two-particle interferometers \cite{2PItheory} has provided a
connection between quantum information processing and edge state
transport.

Another important aspect of edge state transport is the
high-frequency properties. The experiment of Gabelli {\it et al}
\cite{Gabelli} confirmed the quantization of the charge relaxation
resistance, predicted in Ref. \cite{BPT}. In a pioneering experiment
F\`eve {\it et al} \cite{Feve} demonstrated that a mesoscopic capacitor
coupled to an edge state can serve as an on-demand source for
electrons and holes, operating at gigahertz frequencies. The
experiment \cite{Feve} was followed by a number of theoretical works
investigating the accuracy of the on-demand source
\cite{singem1,singem2} and {\it e.g.}  particle colliders with two
synchronized sources \cite{Janine1}. The successful realization of the
electronic on-demand source also motivated new work \cite{Janine2} on
entanglement generation on-demand in the quantum Hall regime
\cite{timeent}. A key feature of \cite{Feve} is that
the on-demand source produces a single stream with alternating
electrons and holes; the current has no dc-component, only
ac-components. For quantum information tasks it would be desirable
to have an on-demand source that produces two separate streams, one with
electrons and one with holes. Such a source implemented in edge states
and operating at gigahertz frequencies would also of be interest for
metrological applications.

\begin{figure}[h]
\centerline{\psfig{figure=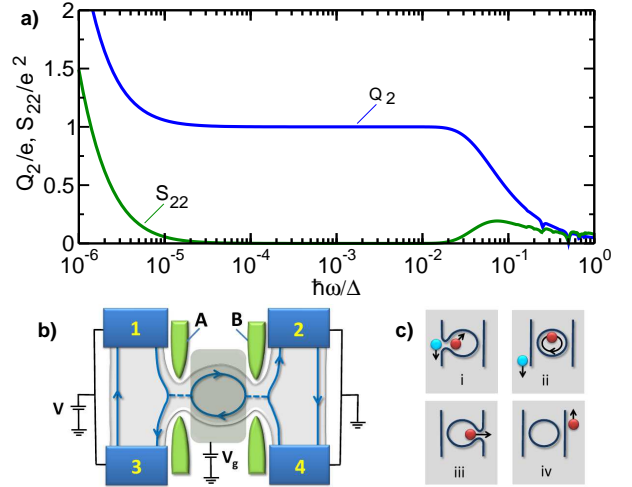,width=8cm}}
\caption{a) Transferred charge $Q_2$ and charge fluctuations
  $S_{22}^{\texttt{neq}}$ per period as a function of frequency with
  $T_A(t),T_B(t)$ shown in Fig. \ref{fig2}a) for $kT \ll \Delta$ (see
  text). b) Schematic of the four terminal turnstile with quantum
  point contacts $A$ and $B$ subjected to time-dependent voltages. The
  top gate (transparent) is kept at a constant voltage $V_g$ and a
  bias $V$ is applied between the two sides. Direction of edge state
  transport shown with arrows. c) Steps in ideal turnstile cycle,
  transporting one hole (blue) to terminal $3$ and one electron (red)
  to $2$.}
\label{fig1}
\end{figure}

In this work we propose such an on-demand source. It comes as a
non-local electron hole turnstile (see Fig. \ref{fig1}) consisting of
a double barrier (DB) formed by two quantum point contacts modulated
periodically in time. A bias voltage is applied between the two sides
of the turnstile, to have one resonant level of the DB in the bias
window. An ideal operation cycle of the turnstile is shown in
Fig. \ref{fig1}: (i) contact $A$ is opened and one electron is
transmitted into the region inside the DB, leaving a hole behind in
the filled stream of electrons continuing towards terminal $3$. (ii)
Contact $A$ closes and subsequently (iii) $B$ opens and the electron
trapped inside the DB is transmitted out through $B$ and (iv)
continues to terminal $2$. Thus, during the cycle exactly one hole and
one electron are emitted into spatially separate terminals.

Since the early turnstile experiments \cite{turnstile1} there has been
large progress in operation speed and accuracy. Recent turnstiles or
single electron pumps have demonstrated operation at gigahertz
frequencies \cite{turnstilegiga} and single parameter pumping
\cite{turnstilesing,Kaschev}. The observed trend with increasing
accuracy at large magnetic fields \cite{bfieldturn} provides
additional motivation for our quantum Hall turnstile.

Our proposal has a number of key features which have not been
addressed together in earlier theoretical \cite{Theory} or
experimental \cite{turnstilesing,turnstilegiga,Kaschev,bfieldturn}
works. First and foremost, the four-terminal edge state geometry gives
spatially separated streams of electrons and holes. This can be
investigated by independent measurements of electron and hole currents
as well as current auto and cross correlations. Second, we do not
restrict ourselves to the tunnel limit but consider operation at
arbitrary contact transparencies, allowing for ideal operation at
higher drive frequencies. Finally, by taking a time-dependent
scattering approach we can analyse both charge currents and
correlations at arbitrary drive frequencies within the same
framework. In particular, we fully account for the fluctuations caused
by the drive of the quantum point contacts, found to only marginally
affect the ideal turnstile operation.

In the following we first present the turnstile system and discuss the
time-dependent current and the charge transferred per cycle in
different driving frequency regimes.  Thereafter the fluctuations of
the charge transfer are investigated.  We consider a DB turnstile
implemented in a four terminal conductor in the quantum Hall regime,
see Fig. \ref{fig1}. The terminals $1,3$ are biased at $eV$
while $2,4$ are grounded. Transport takes place along a single
spin-polarized edge state. Scattering between the edges occurs at the
two quantum point contacts $A$ and $B$. The contacts $A,B$ are created
by electrostatic gates subjected to time-periodic voltages
$V_A(t)=V_{A}^{\texttt{dc}}-V_{A}^{\texttt{ac}}\sin(\omega t)$ and
$V_B(t)=V_{B}^{\texttt{dc}}+V_{B}^{\texttt{ac}}\sin(\omega t)$, $\pi$
out of phase for optimal turnstile operation and with a period ${\mathcal
  T}=2\pi/\omega$.

The transport through the system is conveniently described within the
Floquet scattering approach \cite{buttpump}, applied to a DB-system in
Refs. \cite{buttpump,butttime} with the focus on the quantum pumping
effect.  The time-dependent current flowing into terminal $2$ is
naturally parted into two components,
$I_{2}(t)=I_2^{\texttt{bias}}(t)+I_2^{\texttt{pump}}(t)$. The current
\begin{equation}
I_2^{\texttt{bias}}(t)=\frac{e}{h}\int dE |t_{21}(t,E)|^2[f_V(E)-f_0(E)]
\label{curreq}
\end{equation}
where $f_V(E)$ and $f_0(E)$ are the Fermi distributions of the biased
and grounded terminals respectively. In the absence of an applied bias
$I_2^{\texttt{bias}}(t)$ is thus zero. The dynamical scattering
amplitude \cite{butttime}
$t_{21}(t,E)=t_B(t)\sum_{q=0}^{\infty}e^{i(2q+1)\phi(E)}
L_{q}(t)t_A(t-[2q+1]\tau)$, with $L_q(t)=\prod_{p=1}^q
r_{A}(t-[2p-1]\tau)r_B(t-2p\tau)$ for $q\geq 1$ and $1$ for $q=0$, is
the total amplitude for an electron injected from terminal $1$ at
energy $E$ to be emitted into terminal $2$ at a later time $t$. Here
$\tau=L/v_D$ is the time of flight along the edge from $A$ to $B$ (and
$B$ to $A$), with $v_D$ the drift velocity and $L$ the length. The
phase $\phi(E)=\phi_0+\pi E/\Delta$ where $\Delta=\pi\hbar v_D /L$ the
resonant level spacing in the DB and $\phi_0$ a constant phase,
controlled by the top-gate potential $V_g$, determining the level
positions. The component $I^{\texttt{pump}}_2(t)$ is the pumped
current, independent on bias. It is found to be negligibly small
compared to $I^{\texttt{bias}}_2(t)$ for $\omega \ll \Delta$, with
zero dc-component for all $\omega$, and is only discussed in the
context of the noise below.  The current at terminal $3$ is found
similarly, with $I_3^{\texttt{bias}}=-I_2^{\texttt{bias}}(t+{\mathcal
  T}/2)$ and the transferred charge per cycle is
$Q_2=-Q_3=\int_0^{\mathcal T}I_2(t)dt$.

\begin{figure}[h]
\centerline{\psfig{figure=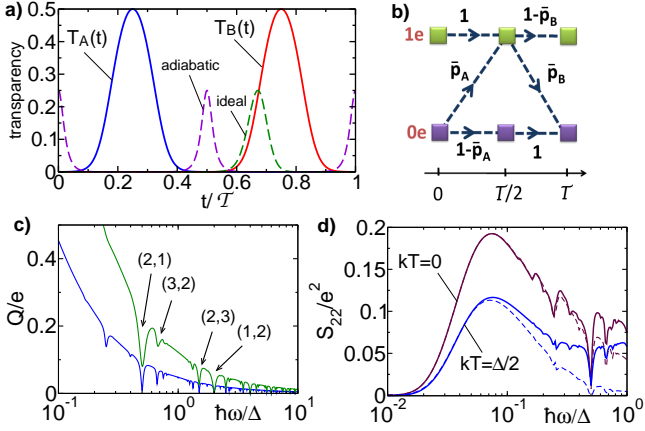,width=8.5cm}}
\caption{a) Transparency $T_A(t)$ and $T_B(t)$ for saddle point
  potential parameters
  $V_{A/B}^{\texttt{dc}}-V_{A/B}^1=V_{A/B}^{\texttt{ac}}=10V_{A/B}^0$. The
  time dependent current $I_2(t)$ (arb. units) in the adiabatic and
  ideal turnstile regimes are shown (dashed lines). b) Illustration of
  the discrete time model of Eq. (\ref{Feq}) with probabilities $\bar
  p_A,\bar p_B$ and directions for transfer between states with 0 and
  1 electrons in the DB-region shown. c) Charge $Q_{2}$ for high
  frequencies, displaying dips described by Eq. (\ref{dipeq}). Number
  of (laps,cycles) shown for four dips. Saddle point parameters as in
  a) (lower curve) and
  $V_{A/B}^{\texttt{dc}}-V_{A/B}^1=4V_{A/B}^{\texttt{ac}}/3=6V_{A/B}^0$
  (upper curve). d) Correlations $S_{22}^{\texttt{neq}}$ with (solid)
  without and (dashed) pumping contribution for high frequencies,
  $kT=0,\Delta/2$ and saddle point parameters as in a).}
\label{fig2}
\end{figure}

The point contact scattering amplitudes $t_{A/B},r_{A/B}$ are taken
energy independent on the scale $\mbox{max}\{kT,eV,\hbar\omega\}$,
with $T$ the temperature.  Motivated by the successful modelling in
\cite{Gabelli}, we describe the contacts A,B with saddle point
potentials \cite{saddlepoint}. The time dependent scattering
amplitudes are $t_{A/B}(t)=i\sqrt{T_{A/B}(t)}$ and
$r_{A/B}(t)=\sqrt{1-T_{A/B}(t)}$ where
$T_{A/B}(t)=(1+\exp[(V_{A/B}(t)-V_{A/B}^1)/V_{A/B}^0])^{-1}$ with
$V_{A/B}^{0/1}$ properties of the potential. Throughout the paper it
is assumed that the product $T_A(t)T_B(t)\ll 1$, a typical driving
scheme is shown in Fig. \ref{fig2}a. The top-gate suppresses charging
effects \cite{Gabelli,Feve}, supporting our non-interacting
approximation.

In the rest of the paper we consider the case with $eV=\Delta$ giving
one DB-level inside the bias window, optimal for the ideal turnstile
operation shown in Fig. \ref{fig1}.  We can then perform the energy
integral in Eq. (\ref{curreq}) giving
\begin{eqnarray}
I_2(t)&=&(\Delta e/h) T_B(t)F(t-\tau), \nonumber \\
F(t)&=&T_A(t)+R_A(t)R_B(t-\tau)F(t-2\tau).
\label{rec}
\end{eqnarray}
Quite remarkably, the current $I_2(t)$ depends only on the scattering
probabilities $T_{A/B}(t)=1-R_{A/B}(t)$ of the contacts $A/B$ at times
earlier than $t$. The result is independent on temperature and holds
for arbitrary driving frequency. The recursively defined $0 \leq F(t)
\leq 1$ is the probability that an electron injected in the bias
window from terminal $1$ at a time $t-2n\tau$ ($n\geq 0$ integer) is
propagating away from $A$ towards $B$ at time $t$.

In the {\it adiabatic} transport regime, the dwell time of the
particles in the DB is much shorter than the drive period
$\mathcal{T}$. The maximum dwell time for particles injected in the
bias window is $\sim \hbar/(\Delta \mbox{min}[T_A(t)+T_B(t)])$, the
inverse of the minimum resonant level width (taken over one
period). Thus, at frequencies $\omega \ll \Delta
\mbox{min}[T_A(t)+T_B(t)]/\hbar$ the transport is adiabatic. The current is found by taking $\tau
\rightarrow 0$ in Eq. (\ref{rec}), giving
\begin{eqnarray}
I_2^{\texttt{ad}}(t)=(e \Delta/h)T_B(t)T_A(t)/[1-R_{A}(t)R_{B}(t)].
\label{adcurr}
\end{eqnarray}
This is simply the instantaneous DB current. Importantly, the
corresponding transferred charge per period $Q_2^{\texttt{ad}} \gg e$
(see Fig. \ref{fig2}), {\it i.e.} many particles traverse the DB
during one period.  From Eq. (\ref{adcurr}) and Fig. \ref{fig2} it is
clear $Q_2^{\texttt{ad}}\propto 1/\omega$ and that
$I_2^{\texttt{ad}}(t)$ flows around times when $T_{A}(t)T_{B}(t)$ is
maximal. Consequently, for a driving where contacts $A$ and $B$ are
never both open at the same time there is no adiabatic current flow,
or equivalently the adiabatic frequency limit $\Delta
\mbox{min}[T_A(t)+T_B(t)]/\hbar \rightarrow 0$.

From this reasoning it follows equally that for frequencies in the
{\it non-adiabatic} regime, $\omega \gg \Delta
\mbox{min}[T_A(t)+T_B(t)]/\hbar$, we can neglect the current flow
during times when both contacts are open. This leads to the standard
physical picture in terms of charging and discharging of the
DB-region: for the cycle $0<t<{\mathcal T}$ (mod ${\mathcal T}$), i)
at times $0<t<{\mathcal T}/2$ contact B closed and charge is flowing
into the DB-region through $A$.  ii) at times ${\mathcal
  T}/2<t<{\mathcal T}$ contact $A$ is closed and charge is flowing out
through $B$.

Focusing first on non-adiabatic frequencies much smaller than the
level spacing, $\hbar \omega \ll \Delta$, the charge density inside
the DB-region is uniform and a calculation of the charge in the
DB-region, injected in the bias window, gives $Q(t)=eF(t)$. Thus,
$F(t)$ is just the probability to find an electron inside the DB. The
time development of the charge is found from Eq. (\ref{rec}),
\begin{equation}
Q(t)=\left\{\begin{array}{cc} p_A(t)+[1-p_A(t)]Q(0) & \mbox{charging} \\  \left[1-p_B(t)\right]Q({\mathcal T}/2) & \mbox{discharging} \end{array}\right.
\label{Feq}
\end{equation}
where {\it e.g.} $p_{B}(t)=1-\prod_{p=0}^{P_{B}}R_{B}(t-2\tau p)$ is
the probability that an electron inside the DB at time ${\mathcal
  T}/2$ has been transmitted out through contact $B$ at time $t$, with
$P_{B}=\mbox{int}[(t-{\mathcal T}/2)/(2\tau)]$. $p_A(t)$ and $P_A$ are
given analogously. The charge at the opening/closing is $Q({\mathcal
  T})=Q(0)=\bar p_A(1-\bar p_B)/(\bar p_A+\bar p_B-\bar p_A\bar p_B)$
and $Q({\mathcal T}/2)=Q(0)/(1-\bar p_B)$ where $\bar
p_A=p_A({\mathcal T}/2)$, $\bar p_B=p_B({\mathcal T})$. The current
$I_2(t)=(\Delta/h)T_B(t)Q(t)$ is shown in Fig. {\ref{fig2}.

  For times $t$ not close to the opening times of $A$ and $B$,
  i.e. $P_{A},P_{B}\gg 1$, we can write {\it e.g.}
  $1-p_{B}(t)=e^{\sum_{p=0}^{P_{B}}\ln[R_{B}(t-2\tau p)]}\approx e^{
    (1/2\tau)\int_{{\mathcal T}/2}^t\ln[R_{B}(t')]dt'}$. It is
  instructive to compare this with the charging and discharging of a
  classical RC-circuit with a capacitance $C$ and a slowly
  time-varying resistance ${\mathcal R}(t)$, for which
  $e^{-\int_{{\mathcal T}/2}^t [C{\mathcal R}(t')]^{-1}dt'}$
  corresponds to $1-p_{B}(t)$. This gives a capacitance $C=e^2/\Delta$
  and a resistance ${\mathcal R}(t)=(h/e^2)/\ln[1/R_{B}(t)]$,
  providing a turnstile analogy of the models for the on-demand source
  discussed in \cite{Feve,singem1}.

  The transferred charge per cycle, $Q({\mathcal T}/2)-Q({\mathcal T})$, is
\begin{equation}
Q_2=-Q_3=e\bar p_A \bar p_B/[\bar p_A+ \bar p_B-\bar p_A\bar p_B].
\end{equation}
This gives that for $\omega \ll
\omega_A^{\texttt{max}},\omega_B^{\texttt{max}}$ with $\hbar
\omega_{A/B}^{\texttt{max}}=\Delta \mbox{min}\{1,\int_0^{{\mathcal T}}
(dt/{\mathcal T})\ln[1/R_{A/B}(t)]\}$, we have $\bar p_A,\bar p_B=1$
and $Q_2=-Q_3=e$ {\it i.e.}  exactly one electron and one hole are
transferred. Taken together, this yields a frequency interval $ \Delta
\mbox{min}[T_A(t)+T_B(t)]/\hbar \ll \omega \ll
\omega_{A}^{\texttt{max}},\omega_{B}^{\texttt{max}}$ for the ideal
turnstile cycle shown in Fig. \ref{fig1}. For higher frequencies
electrons do not have time to completely charge or discharge the
DB-region and $Q_2<e$.

Importantly, for tunnelling contacts $T_A(t),T_B(t) \ll 1$ and $\omega
\ll \Delta/\hbar$ we can directly expand $F(t-2\tau)=F(t)-2\tau
dF(t)/dt$ in Eq. (\ref{rec}) and arrive at
\begin{equation}
dP_1/dt=-\Gamma_A(t)P_1(t)+\Gamma_B(t)P_0(t)
\label{ME}
\end{equation}
where $P_1(t)=F(t)=1-P_0(t)$ and
$\Gamma_{A/B}(t)=T_{A/B}(t)\Delta/h$. This is a master equation
with time dependent tunnelling rates, investigated in
e.g. \cite{Flensberg,MEFCS,Kaschev}.

At frequencies $\omega \sim \Delta/\hbar$ the expression in
Eqs. (\ref{Feq}) and (\ref{ME}) break down and transport through
higher/lower lying resonances become visible, manifested as sharp dips
in the transferred charge as a function of frequency, see
Fig. \ref{fig2}c. The most pronounced set of dips, at frequencies
\begin{equation}
\hbar \omega=\Delta(2n+1\pm 1/m) 
\label{dipeq}
\end{equation}
results from electrons, which after being injected at $A$ at maximal
$T_A(t)$, circulate around the DB-region $m$ times during $m(2n+1) \pm
1$ periods before escaping back out at $A$ [at maximal $T_A(t)$], not
transferring any charge.

For a long measurement time $t_0=N{\mathcal T}, N \gg 1$, to
characterize the accuracy of the turnstile it is important to
investigate not only the average charge transferred per cycle,
$Q_2=(1/N)\int_0^{t_0} dtI_2(t)$, but also the fluctuations
\cite{buttpump,butttime,noisetheory}, experimentally accessible via
current correlations \cite{noise}. To this end we first write the
current $I_2(t)=\sum_q i_{2,q}\exp(iq\omega t)$, with
$i_{2,q}=(e/h)\int dE j_{2,q}(E)$ and
$j_{2,q}(E)=j_{2,q}^{\texttt{bias}}(E)+j_{2,q}^{\texttt{pump}}(E)$
where
\begin{eqnarray}
j_{2,q}^{\texttt{pump}}(E)&=&\sum_{n}\left[T_{21}^{q,n}(E)+T_{24}^{q,n}(E)\right][f_0(E_n)-f_0(E)] \nonumber \\
j_{2,q}^{\texttt{bias}}(E)&=&\sum_{n}T_{21}^{q,n}(E)[f_V(E_n)-f_0(E_n)].
\label{currcomp}
\end{eqnarray}
Here $E_n=E+n\hbar\omega$,
$T_{2\alpha}^{q,n}(E)=t_{2\alpha}^*(E,E_n)t_{2\alpha}(E_{-q},E_n)$,
$\alpha=1,4$ and $t_{2\alpha}(E_m,E)=\int_0^{\mathcal T} (dt/{\mathcal
  T})e^{im\omega t}t_{2\alpha}(t,E)$ with $t_{21}(t,E)$ given above
and $t_{24}(t,E)=r_B(t)+t_B(t)\sum_{q=0}^{\infty}
e^{i2(q+1)\phi(E)}L_q(t)r_A(t-[2q+1]\tau)t_B(t-2[q+1]\tau)$. The
current at terminal $3$ is found similarly.

The auto-correlations of transferred charge at terminal $2$ is
$S_{22}=(1/N)\int_0^{t_0}\int_0^{t_0}dt dt'\langle \Delta I_2(t)\Delta
I_2(t')\rangle$ where $\Delta I_2(t)$ is the current fluctuations
\cite{Buttrev}. Calculations following Ref. \cite{buttpump} give
$S_{22}=S_{22}^{\texttt{neq}}+S^{\texttt{th}}$ with
\begin{equation}
S^{\texttt{neq}}_{22}=\frac{{\mathcal T}e^2}{h}\int dE\left[j_{2,0}[1-2f_0(E)]-\sum_q|j_{2,q}|^2\right]
\label{noise}
\end{equation}
and $S^{\texttt{th}}=2{\mathcal T}(e^2/h)kT$ the thermal noise in the
absence of both drive and bias. The auto correlator $S_{33}$ and the
cross correlators $S_{32}=S_{23}$ are found similarly.

We first consider the correlations at $\hbar\omega,kT \ll \Delta$.  In
this regime the fluctuations are minimized for DB-levels at energies
$\Delta(n+1/2)$; one level in the middle of the bias window.  In
particular we find that the pumping components
$j_{2,q}^{\texttt{pump}}(E)$ contribute negligibly to the correlations
(dc-component $i_{2,0}^{\texttt{pump}}=0$) and hence
$S^{\texttt{neq}}_{22}=S^{\texttt{neq}}_{33}\equiv
S^{\texttt{bias}}=({\mathcal T}e^2/h)\int
dE[j_{2,0}^{\texttt{bias}}-\sum_q|j_{2,q}^{\texttt{bias}}|^2]$.
Importantly, the total cross-correlator $S_{23}=-S^{\texttt{bias}}$,
independent on equilibrium thermal fluctuations. This allows for an
independent investigation of the turnstile accuracy.

The reason for the negligible pumping noise can be understood as
follows: The term $j_{2,q}^{\texttt{pump}}(E)$ describes creation of
electron-hole pairs close to the Fermi energy [from
$f_0(E_n)-f_0(E)$]. DB-levels at $\Delta(n+1/2)$ imply completely
off-resonance Fermi energy transport, strongly suppressing the
electron-hole pair creation. Formally, the terms in Eq. (\ref{noise})
containing $j_{2,q}^{\texttt{pump}}$ are found to be of order $(\hbar
\omega/\Delta)\int_0^{{\mathcal T}}(dt/{\mathcal T})T_A(t)T_B(t)$
smaller than terms from $j_{2,q}^{\texttt{bias}}$. This is supported
by the numerics in Fig. \ref{fig2}.

In the {\it adiabatic} regime the correlations are found by
inserting the frozen scattering amplitudes into the expression for
$S^{\texttt{bias}}$, giving the time integral over one period 
of the instantaneous DB shot-noise \cite{Buttrev}. In the {\it non-adiabatic} 
regime the full distribution of the transferred charge can be found from
Eq. (\ref{Feq}), describing the time-evolution of the probability
$F(t)=Q(t)/e$ to have one electron inside the DB. The possible processes
taking the DB between charge states with 0 and 1 electrons at times
$t=0$ and $t={\mathcal T}/2$ (mod ${\mathcal T}$) are
shown in Fig. \ref{fig2}. The full counting statistics for the
charge transfer, described by a time-discrete master
equation, is known \cite{DBFCS,MEFCS}. The generating function for the
probability distribution is 
\begin{equation}
\xi(\lambda_2,\lambda_3)=N\ln\left[h+\sqrt{h^2+(1-\bar p_A)(1-\bar p_B)}\right]
\end{equation}
where $h=1-(\bar p_A+\bar p_B-\bar p_A\bar
p_B\mbox{exp}[i(\lambda_2-\lambda_3)])/2$ and $\lambda_2,\lambda_3$ the
counting fields. The cumulants are obtained by taking succesive
derivatives of $\xi(\lambda_2,\lambda_3)$ with respect to $\lambda_2,\lambda_3$. Here we focus on the he second cumulant, 
\begin{equation}
S^{\texttt{bias}}=\frac{e^2\Delta}{h}\frac{\bar p_A\bar p_B[\bar p_A^2(1-\bar p_B)+\bar p_B^2(1-\bar p_A)]}{(\bar p_A+\bar p_B -\bar p_A\bar p_B)^3}.
\label{FCSnoise}
\end{equation}
In the ideal regime, $\bar p_A=\bar p_B=1$, the
noise is zero. For higher frequencies the noise increases due to the stochastic 
charging and discharging of the DB-region, see Fig. \ref{fig2}. For large frequencies $\omega \sim \Delta$ both the components $j_{2,q}^{\texttt{pump}}$ and
$j_{2,q}^{\texttt{bias}}$ contribute to the correlations. The correlations are
evaluated numerically, shown in Fig. \ref{fig2}. 

In conclusion we have analysed a mesoscopic turnstile implemented in a
double barrier system in the quantum Hall regime. At ideal operation
the turnstile produces one electron and one hole at different
locations per driving cycle. The noise due to the driving is found to
be negligibly small at frequencies for ideal operation.

We acknowledge discussions with M. B\"uttiker, M. Moskalets,
J. Splettstoesser, C. Flindt, M. Albert, G. F\`eve and A. Wacker and
support from the Swedish VR.

\end{document}